\documentclass[prl,aps,superscriptaddress,floatfix,twocolumn]{revtex4-2}
\usepackage{graphicx,graphics,psfrag,amsmath,calc}
\usepackage{color,comment}
\usepackage{amsfonts}
\usepackage{tikz}
\usepackage[normalem]{ulem} 
\usepackage[ruled,linesnumbered]{algorithm2e}
\usepackage{natbib}
\usepackage[bookmarks=true,colorlinks=true,urlcolor=blue,citecolor=blue,linkcolor=blue]{hyperref}
\usepackage[]{cleveref}
\usepackage{titlesec}

\definecolor{vermillion}{rgb}{0.86, 0.18, 0.01}

\definecolor{some-color}{rgb}{0.75, 0.25, 0.75}

\def\sinc{\text{sinc}}

\begin{document}
\title{Quantum simulation of $\phi^4$ theories in qudit systems}

\author{
Doga Murat Kurkcuoglu
}
\affiliation{Fermi National Accelerator Laboratory, Batavia,  IL, 60510, USA}
\affiliation{Superconducting Quantum Materials and Systems Center (SQMS),
Fermi National Accelerator Laboratory, Batavia, IL 60510, USA
}

\author{
M. Sohaib Alam
}
\affiliation{Rigetti Computing, Berkeley, CA, 94701, USA}
\affiliation{Superconducting Quantum Materials and Systems Center (SQMS),
Fermi National Accelerator Laboratory, Batavia, IL 60510, USA
}

\author{
Joshua Adam Job
}
\affiliation{Lockheed Martin Advanced Technology Center, Sunnyvale, CA, 94089}

\author{
Andy C.~Y.~Li
}
\affiliation{Fermi National Accelerator Laboratory, Batavia,  IL, 60510, USA}
\affiliation{Superconducting Quantum Materials and Systems Center (SQMS),
Fermi National Accelerator Laboratory, Batavia, IL 60510, USA
}

\author{
Alexandru Macridin
}
\affiliation{Fermi National Accelerator Laboratory, Batavia,  IL, 60510, USA}
\affiliation{Superconducting Quantum Materials and Systems Center (SQMS),
Fermi National Accelerator Laboratory, Batavia, IL 60510, USA
}

\author{
Gabriel N. Perdue
}
\affiliation{Fermi National Accelerator Laboratory, Batavia,  IL, 60510, USA}
\affiliation{Superconducting Quantum Materials and Systems Center (SQMS),
Fermi National Accelerator Laboratory, Batavia, IL 60510, USA
}

\author{
Stephen Providence
}
\affiliation{Coppin State University, Baltimore, MD, 21216, USA}

\date{\today}

\begin{abstract}
We discuss the implementation of quantum algorithms for lattice $\Phi^4$ theory on a circuit quantum electrodynamics (cQED) system.
The field is represented on qudits using a discretized field amplitude basis.
The main advantage of a qudit system is that its multi-level characteristics allows the field interaction to be implemented with only diagonal single-qudit gates.
Considering  the set of universal gates formed by the single-qudit phase gate and the displacement gate, we address initial state preparation and single-qudit gate synthesis with variational methods. 

\end{abstract}

\maketitle

\section{Introduction}
\label{sec:intro}


Bosonic fields are ubiquitous in physics, from  particle physics models such as the Higgs boson \cite{higgs1964broken}, gauge bosons such as photons and gluons \cite{fradkin1979phase} or Skyrme model \cite{skyrme1994non}  to effective field models that describe collective 
excitations in condensed matter physics such as phonons, magnons, plasmons, etc.
The simulation of real time evolution of quantum fields is  difficult to address analytically or with classical simulations when the dimensionality of the problem is big.
For real-time classical simulations of scalar fields are limited to small systems since the memory requirement increases exponentially with system size.
This computational difficulty has inspired proposals to study field theory simulations on qubit-based quantum computers \cite{somma2003quantum, Jordan2012, macridin2018digital, li2021phi4, barata2021single, klco2019digitization,  martin2021long, klco2020systematically, klco2020fixed, muschik2017u, klco2021geometric, farrelly2020discretizing}. 
Another path forward for studying the dynamics of field theories is to utilize cold atoms in optical lattices and simulate the field in an actual quantum environment \cite{Banuls2020, surace2020lattice, mil2020scalable}. 
Relatively little work has been done on qudit (N-level) systems, although recently qudit simulations of a 1+1 QED model and lattice gauge theories were discussed in~\cite{gustafson2021prospects, jordan2011quant, zache2021achieving, luo2020gauge, padmanabhan2014more, hashimoto2017time, bacon2006efficient}. 
In addition to the simulation of field theories in discrete lattice, simulation of continuous variables is also possible \cite{marshall2015quantum, bartlett2002efficient}
Here we propose to use high-dimensional qudits ($N_b \ge 10$), where $N_b$ is the number of discretization points in a qudit,  for the simulation of scalar field dynamics. 

The purpose of this work is to set out the necessary ingredients for real time simulation of scalar fields on qudit based platforms, including initial state preparation and gate synthesis for the Trotter steps.
Recent advances in cQED systems make the platform an attractive candidate for field theory simulations \cite{blais2004cavity}. 
In cQED systems, photon levels can be encoded and manipulated for qudit based quantum computation. 
The number of levels in a qudit are not restricted to two as is the case with qubit based platforms; thus the algorithms, gates, and state preparation for qudits require a separate discussion from their qubit counterparts. 
An advantage of high-dimensional qudit quantum simulations is that the field at every lattice site  can be encoded in only a single qudit, unlike the qubit simulations where
the local field is represented on many qubits.
Single-qudit encoding of local fields also implies single-qudit gates for the implementation
of local interactions. The interaction implementation in our model takes
advantage of one of the most attractive experimental capability of cQED systems, namely
the ability to easy manipulate the phase of each photon number state~\cite{heeres2015cavity}. 
This experimental technique makes the field theory simulation rather straightforward in qudit based quantum computation. 
We discuss a field theory simulation algorithm with $\Phi^4$ type interaction term in qudit based platforms and we demonstrate a short-time evolution example, simulated on a classical computer. 

The paper is organized as follows: First we define the theory and Hamiltonian.
We discuss the discretization of the field and expansion in harmonic oscillator basis. 
We then present two sections --- one focused on a single-qudit computer, and one on a mult-qudit system.
In the single qudit section, we discuss state preparation and gate preparation with variational algorithms, and how to find the ground state of a field with nonlinearities present. 
In the multiple qudits section, we discuss how a field might be modeled in entangled cavities. 
In a final section we show the simulation algorithm for the full Hamiltonian. 

\subsection{Definition of the theory}
\label{sec:theorydef}

We consider the $\Phi^4$ scalar field theory, defined by the Lagrangian \cite{wilson1969non}:
\begin{eqnarray}
\mathcal{L} 
=
\frac{1}{2}
\left(\partial_0 \hat{\Phi}\right)^2
-
\frac{1}{2}
\left(\nabla \hat{\Phi}\right)^2
-
\frac{1}{2}
m_0^2
\hat{\Phi}^2
-\frac{\lambda}{4!}\hat{\Phi}^4,
\label{eq:lagrangian}
\end{eqnarray}
where 
$
\hat{\Phi} \equiv \hat{\Phi}({\bf r}, t)
$
is shorthand notation for a scalar field with eigenvalue $\Phi$, $\hat{\Phi}({\bf r}, t)|\Phi \rangle = \Phi({\bf r}, t) |\Phi \rangle$  
that is dependent on the position vector ${\bf r} = (r_1, r_2, r_3)$ and time $t$, 
$
\partial_0 \equiv \partial /\partial t
$
is the time derivative, $\hbar = 1$, and we use the $(+,-,-,-)$ sign convention for the Minkowski metric. 
In this work, we show time simulation for a 1+1 dimensional field theory, i.e. one spatial and one time degree of freedom.
However, extension into higher dimensions is straightforward. 
The time simulation of a field will be realized with consecutive application of selected qudit gates such that the amplitudes of the Fock states in a qudit are manipulated. 

The corresponding Hamiltonian density is obtained via a Legendre transformation of the Lagrangian, 
\begin{equation}
\mathcal{H}
=
\left(\partial_0 \hat{\Phi}\right) \hat{\pi} - \mathcal{L},
\end{equation}
where $\hat{\pi} = \partial_0 \hat{\Phi}$ is the canonical momentum that satisfies the commutation relation $\left[\hat{\Phi}({\bf r},t), \hat{\pi}({\bf r'},t')\right]=i\delta({\bf r}-{\bf r'})\delta(t-t')$. 
The Hamiltonian density for the $\Phi^4$ theory is
\begin{eqnarray}
\label{eq:phi4thhamilton}
\mathcal{H}
=
\frac{1}{2}\hat{\pi}^2 + \frac{1}{2}
\left(\nabla \hat{\Phi}\right)^2 + \frac{1}{2}m_0^2\hat{\Phi}^2 + \frac{\lambda}{4!}\hat{\Phi}^4.
\end{eqnarray}
In order to do numerical simulations the continuous field is discretized on a lattice, $\Phi \rightarrow \Phi_j(t)$ where $j$ is a lattice site index.
The lattice Hamiltonian reads
\begin{multline}
\mathcal{H}_d
=
a^d\sum \limits_j\left[\frac{1}{2}\hat{\pi}_j^2+\frac{1}{2}m_0^2\hat{\Phi}_j^2+ \right. \\ \left. \frac{\lambda}{4!}\hat{\Phi}_j^4 + \frac{1}{2a^2}\sum \limits_{e\neq j}^d\left(\hat{\Phi}_{j+e}-\hat{\Phi}_j\right)^2\right],
\end{multline}
where $d$ is the spatial dimension (here $d=1$), $a$ is the lattice constant and $e$ is the index for the nearest-neighbor site.
The commutation relation for the discretized field is $[\hat{\Phi}_j, \hat{\pi}_k] = ia^{-d}\delta_{j,k}$, where $\delta_{i,j}$ is the Kronecker delta. 
For clarity we scale the fields such that $\hat{\phi}_j = a^{\frac{d-1}{2}}\hat{\Phi}_j$, $\hat{\Pi}_j = a^{\frac{d+1}{2}}\hat{\pi}_j$, the bare mass $\mu^2 = m_0^2a^2$, and the dimensionless bare coupling constant $g = \lambda a^{3-d}$.
The renormalized Hamiltonian is then:
\begin{multline}
\label{eq:scaledHamiltonian}
\bar{\mathcal{H}} = \sum \limits_j \left[\frac{\hat{\Pi}_j^2}{2} + \frac{1}{2}\left(\mu^2 + 2d\right)\hat{\phi}_j^2 - \right. \\ \left. \sum \limits_{e=1}^{d}\hat{\phi}_j\hat{\phi}_{j+e} + \frac{g}{4!}\hat{\phi}_j^4\right],
\end{multline}
where $\bar{\mathcal{H}} = a\mathcal{H}_d $. 

cQED systems are QED systems with artificial atoms (superconducting qubits) which is coupled to one \cite{blais2004cavity} or multiple cavity modes \cite{chakram2021seamless}.
In cQED systems, the EM fields inside a cavity can be manipulated via the transmon or by directly applying a control signal to the EM field. 
The resonator in which the TEM fields oscillate may be two-dimensional or three-dimensional. 
3D cQED systems are well-suited to time-simulate a field $\phi$ due to their versatility, the ability to manipulate cavity modes \cite{chakram2020multimode, chakram2021seamless}, and longer coherence times \cite{romanenko2020three, siddiqi2021engineering}. 

A qudit may support more than two levels, unlike a (logical) qubit. 
The Fock states in a cavity may be used to represent the fields. 
Thus, we will refer to the Fock states in a cavity as the logical subspace of a qudit. 
These states allow us to represent one discretized $\phi$ field using a single qudit.

The manipulation of the amplitudes of the Fock states in a cavity can be made via selective phase gates. 
This requires the phase gates to be proportional to the photon number of the cavity ($n$). 
The phase that each state gains can be engineered to be linearly proportional to the photon number $n$, or the photon number to any arbitrary power $k$ of the photon number $n^r$.
This may be engineered by driving the transmon with a signal frequency that is dependent on the photon number \cite{heeres2015cavity}. 
The qudit phase gate is known as the selective number of arbitrary photon (SNAP) gate.
This offers a new and convenient platform for the simulation of field theories in cavity systems. 

The Hamiltonian~\eqref{eq:scaledHamiltonian} describes a set of coupled self-interacting harmonic oscillators. The Hilbert space of the system 
\begin{align}
    {\cal{H}}=\prod_{j=0}^{L-1} \otimes {\cal{H}}_j
\end{align}
is a product of local Hilbert spaces ${\cal{H}}_j$
where $j$ is the lattice site label. 
A possible basis choice for the local Hilbert space is the field boson occupation number,
\begin{align}
    |b_n\rangle_j=\frac{1}{\sqrt{n!}}\left(b_j^{\dagger}\right)^n|0\rangle_j
\end{align}
where the field boson creation operator is $b_j^{\dagger}=\left(\sqrt{m}\hat{\phi}_j-i\hat{\Pi}_j/\sqrt{m}\right)/\sqrt{2}$. For numerical simulations the local Hilbert space is truncated by introducing a boson occupation cutoff $N_b$. 
The truncated local Hilbert space spanned by $\{|b_n\rangle_j\}_{n=\overline{0,N_b-1}}$  can be 
represented in a field amplitude discretized 
basis $\{|\phi_\nu\rangle_j\}_{\nu=\overline{0,N_{\mathrm{cut}}-1}}$. The dimension $N_{\mathrm{cut}}$ of the field amplitude basis is larger than $N_b$.
 The representation accuracy increases exponentially with increasing $N_{\mathrm{cut}}$. For example, a choice $N_{\mathrm{cut}}=2N_b$ ensures a $10^{-4}$ accuracy \cite{macridin2021bosonic}.
We map the discretized field amplitude vectors on qudit 
states. The local field operator act on these states as
\begin{align}
    \hat{\phi}_j |\phi_\nu\rangle_j= \left(\nu-\frac{N_{\mathrm{cut}}-1}{2}\right) \Delta_{\phi} |\phi_\nu\rangle_j ~~\text{ with}~~\nu=\overline{0,N_{\mathrm{cut}}-1}
\end{align}
where the discretization field amplitude interval is
$\Delta_{\phi}=\sqrt{\frac{2 \pi}{N_{\mathrm{cut}} m}}$. 
The local conjugate field operator can be written as 
\begin{align}
    \hat{\Pi}_j =m\mathcal{F}_{N} \hat{\phi}_j \mathcal{F}_{N}^{-1},
\end{align}
where $\mathcal{F}_{N}$ is the single-qudit $N \times N$ discretized Fourier transform.
We expand a single field eigenvector $\left\{|\phi\rangle_j\right\}$ into the Hilbert space of the $j$-th qudit as 
\begin{equation}
|\phi\rangle_j = \sum \limits_{\nu=0}^{N_{\mathrm{cut}}}c_\nu^j(t)|\phi_{\nu}\rangle_j,
\end{equation}
where $N_{\mathrm{cut}}$ is the cutoff for the Hilbert space dimension. 
We then map this qudit basis into harmonic oscillator eigenspace as
\begin{eqnarray}
\langle x | \phi\rangle_j \equiv \phi_j(t) = \sum \limits_{\nu=0}^{N_{\mathrm{cut}}} c_{\nu}^j(t)\varphi_{\nu}(x),
\end{eqnarray}
where $\varphi_{\nu}(x) \equiv \langle x | \phi_{\nu}\rangle$ are the scaled harmonic oscillator (HO) eigenfunctions 
\begin{eqnarray}
\varphi_{\nu}(x) = \frac{1}{\pi^{1/4}\sqrt{2^{\nu} {\nu}!}}e^{-\frac{1}{2}x^2}H_{\nu}(x),
\end{eqnarray}
where $x$ is the HO displacement scaled by $1/\sqrt{\hbar/(m\omega)}$
and $H_{\nu}(x)$ are the Hermite polynomials \cite{macridin2021bosonic}. 
In this notation, the eigenvalue of the single scaled field $\hat{\phi}$ at discretized $x_{\nu}$ becomes $\langle x_{\nu} | \hat{\phi} | \phi_{\nu}\rangle = x_{\nu}\langle x_{\nu} | \phi_{\nu} \rangle$
We use the Fock states in one qudit to discretize the field $\varphi$ in field amplitude basis,
\begin{eqnarray}
\varphi_n(x) = \sum \limits_{\nu=0}^{N_{\mathrm{cut}}-1}\varphi_n(x_{\nu})u_{\nu}(x)+\mathcal{O}(\epsilon),
\end{eqnarray}
where $u_{\nu}(x) = \sinc((x-x_{\nu})/\Delta)$ is the auxiliary function \cite{macridin2018digital}.
The $\epsilon$ is the total error due to the discretization of the eigenfunction $\varphi_n(x_{\nu})$, meaning that $|\varphi_n(x_{\nu})| < \epsilon$ for all $N_{\mathrm{cut}}> {\nu} \geq N_b$.

To study the time-propagation of the field, we Trotterize the Hamiltonian into infinitesimal time-steps $\delta t$,
\begin{eqnarray}
e^{-i\bar{\mathcal{H}}T}= \left(e^{-i\bar{\mathcal{H}}\delta t}\right)^K.
\end{eqnarray}
The HO eigenfunctions are discretized into $N$ states in a qudit.
The gates required for implementing the Trotter steps are, 
\begin{eqnarray}
e^{-i\xi n^r}, e^{-i \xi n m}, \mathcal{F}_{N},
\end{eqnarray}
where $\xi$ is an arbitrary angle that is proportinal to the step length, $n$ is the photon number in one Fock state, $r \geq 1$ is the exponent which is an integer, and $\mathcal{F}_N$ is the $N\times N$ Fourier transform operator, which is an $N\times N$ Hadamard gate with the elements of the $\mathcal{F}_N$ matrix defined as:
\begin{eqnarray}
V_{\mathcal{F}} = \left(\mathcal{F}_N\right)_{l,m} = \frac{1}{\sqrt{N}} e^{i\left[\left(l - N/2\right)\left(m - N/2\right) \right]2\pi/N}.
\end{eqnarray}
The third gate is the coupling term where the photon numbers $n$ and $m$ of two cavities are coupled.
We discuss the multiple qudit case below.

Any arbitrary $N\times N$ unitary gate $U$ may be arbitrarily decomposed into SNAP and displacement gates with an appropriate choice of parameters \cite{krastanov2015universal}. 
We define the truncated displacement and SNAP gates for a single qudit as follows
\begin{eqnarray}
D(\alpha) &=& e^{a \alpha - a^{\dagger}\alpha^*}\\
S_N^{(r)}(\vec{\theta}) &=& \sum \limits_{n=0}^{N-1}|n\rangle\langle n| e^{i\theta_n n^r},
\end{eqnarray}
where $\theta_n$ is an element of the vector $\vec{\theta}$. 
The parameters for SNAP and displacement gates may be found using variational methods. 
Mathematically, the matrix decomposition argument for a single-qudit may be straightforwardly applied to multiple cavities which are coupled to each other and an arbitrary multiqudit gate $U_{(d>1)}$ can also be decomposed into multiqudit displacement and multiqudit SNAP gates.
However, a variational search for the parameters for SNAP and displacement gates for large $N$ values is computationally non-trivial. 
Further, creating conditional SNAP gates for multiple cavity platforms will require a more sophisticated computational and experimental approach.
Thus in this work, the variational approach to engineer qudit gates will be restricted to the single qudit case. 
The parameters that are used to construct single qudit gates are assumed to be useful in the multiqudit gates which are tensor product of these single qudit ones. 
The simulation of $\Phi^4$-th type theory algorithm we presented here can be reduced to single-qudit gates, which describe the evolution associated with local terms of the Hamiltonian, and the conditional SNAP gates are required for the non-local terms, such as $\Phi_j \Phi_{j+e}$. Here we use a variational search to determine the parameters necessary for the single-qudit gates. 
In the next section, we will discuss the state preparation and gate engineering for single qudit problems. 

%
%

\section{Single qudit}

In this section, we discuss state preparation and gate creation for a simulation based on a single qudit. 
The simulation of field theories in qubit systems has been extensively studied over the last two decades \cite{somma2003quantum, trabesinger2012quantum, ladd2010quantum, paulson2020towards, atas20212, Jordan2012, Banuls2020, georgescu2014quantum, macridin2021bosonic}. 
In these simulations, the fields are first encoded in binary form in entangled qubits. 
In order to time-simulate a single field in qubit systems, many one-qubit and two-qubit gates must be consecutively applied to the entangled state.
With qudit SNAP gates, simulating a field can be realized with a single gate. 
The SNAP gates with few Fock states in one qudit can be realized with 0.9 state fidelity \cite{heeres2015cavity}. 
There are also recent efforts to realize multiqudit SNAP gates \cite{chakram2020multimode}.

Let us begin by considering the simplest case, where we want to Trotter-simulate only a single $\phi_j$ field. Although the single field evolution operator alone is not in interest for our Trotter simulations,  
the discussion for single field is instructive before moving onto quadratic or quartic field evolutions since they follow a parallel idea. 
The operator that we need to apply to the qudit state is $e^{-i \beta \phi_j \Delta t}$, with $\beta$ is an unimportant appropriate constant to keep the units consistent. 
Operation on a qudit state requires a gate like $e^{-i \beta (n - N/2)\Delta \delta t}$. 
This is equivalent to a SNAP gate with $e^{-i(\beta\Delta \delta t) n} \equiv e^{-i\xi n}$ and a global phase of $e^{i\beta N/2\Delta \delta t}$ on a single qudit. 
Since we work with $N$ Fock states, the SNAP and displacement gates must be truncated to $N$ states. 
This could create a problem for the displacement gate, where the Fock states beyond $N$ levels are coupled to the first $N$ levels. 
It was shown that if the mean occupation number $\langle n \rangle$ in a Fock state is less than the cutoff photon number $N_{\mathrm{cut}}$, where $\langle n \rangle << N_{\mathrm{cut}}$, the difference between infinite displacement gate and truncated displacement gate is negligible \cite{miranowicz2014phase}.
The truncated annihilation operator $a$ does not satisfy the usual commutation relation but rather $\left[a, a^{\dagger}\right] = 1 - N|N-1\rangle\langle N-1 |$.
The simulation of time evolution for higher order fields such as $\phi_j^2$, $\phi_j^4$ etc. is going to be similar to that of a linear field. 
Consider the evolution of the quadratic term, $(1/2)(\mu^2 + 2d)\phi_j^2$ --- this requires a Trotter operator:
\begin{equation}
\label{eq:quadraticterm}
\begin{aligned}
V_{\phi^2}\equiv e^{-i\frac{1}{2}\left(\mu^2 + 2d\right)\phi_j^2\delta t} 
= 
\prod \limits_{n=0}^{N-1}|n\rangle\langle n|e^{-i\Omega_n(n-N/2)^2}\\
= S_N^{(2)}(-\vec{\Omega})S_N^{(1)}(N\vec{\Omega})S_N^{(0)}(-(N^2/4)\vec{\Omega}),
\end{aligned}
\end{equation}
where $\vec{\Omega} \equiv \left\{\Omega_n\right\}$ is a $N$-vector whose elements are equal to $\Omega_n = (1/2)(\mu^2 + 2d)\Delta^2\delta t$ and $\mu^2$ is the mass term.
When $\mu$ is taken to be an imaginary number, the symmetry breaking phase $\phi \rightarrow -\phi$ can be studied by simply changing the overall sign of the phases of the SNAP gates.

Single-qudit gates may be engineered by means of variational parameters or finding an optimal signal \cite{heeres2017implementing} . 
We construct the required gates and perform state preparation using $S_N^{(k)}(\vec{\theta})$ and $D(\alpha)$ gates by variationally finding the $\theta_n$ and $\alpha$ parameters by minimizing a cost function.
The variational construction of gates involves blocks of single-qudit SNAP and displacement gates \cite{krastanov2015universal} 
$B(\vec{\theta}, \alpha) = D(\alpha)^{\dagger}S_N(\vec{\theta})D(\alpha)$ that are combined to construct a unitary gate,
$U(\vec{\alpha}, \vec{\Theta}) = \Pi_{i=1}^{k} B(\vec{\theta}_k,\alpha_k)$.
Variational optimizaiton is not required for the phase gates, but is employed for gates such as the Fourier transform.
One difficulty in this construction is that for a fixed single qudit state number $N$, the displacement gate excites states beyond $N$ as the creation operator $a^{\dagger}$ couples the adjacent states. 
To manage excitations in the higher and the lower Fock states, we add small number of additional $m$ levels of qudit states at higher Fock states. 
The first $m$ Fock states and the last $m$ Fock states are going to be called the bumper states. The $N$ Fock states in between these bumper states will represent the $\phi_j$ field and they will be called `logical states'.
This means taking the direct sum of logical states $|\psi_l\rangle$ and bumper states $|\psi_b\rangle$,
$
|\psi\rangle =  |\psi_l\rangle \oplus|\psi_b\rangle
$.
In our algorithm, we first prepare the single-qudit state in the cavity ground state, $|\psi\rangle_{t=0} = |0\rangle$, where $|\psi\rangle_{t=0}$ is the initial state. 
Then, we variationally find the parameters of SNAP and displacement gates to have the $c_n(t)$ amplitudes represent a target state in a qudit
\begin{eqnarray}
|\psi\rangle = \sum \limits_{n=0}^{N+m-1}c_n(t) |n\rangle.
\end{eqnarray}

The cost function that we use for state preparation is
\begin{multline}
\mathcal{L}_{\mathrm{state}}
=
\left|\langle \psi| U(\vec{\alpha}, \vec{\Theta})|0 \rangle -1\right|^2 +
\left|\mathcal{P}_m^b U(\vec{\alpha}, \vec{\Theta})|0\rangle \right|^2,
\end{multline}
where $|0\rangle$ is the ground state of the cavity and $|\phi\rangle$ is the target state and 
the parameters that minimizes the cost function are $\vec{\alpha} = (\alpha_1, \alpha_2, ..., \alpha_k)$ and $\Theta = (\vec{\theta}_1, \vec{\theta}_2, ..., \vec{\theta}_k)$. 
The two terms are to make sure that the contribution of the bumper states are minimal. 
The projection matrix $\mathcal{P}_m$ for bumper states is 
\begin{equation}
\mathcal{P}_m^b = \left(
\begin{array}{c|c}
 \textbf{0}_N & 0 \\ \hline
 0 & \textbf{1}_m
\end{array}
\right),
\end{equation}
where $\textbf{1}_m$ is $m\times m$ identity matrix and $\textbf{0}_N$ is the $N\times N$ zero matrix. 

\begin{figure}
\includegraphics[width=0.45\textwidth]{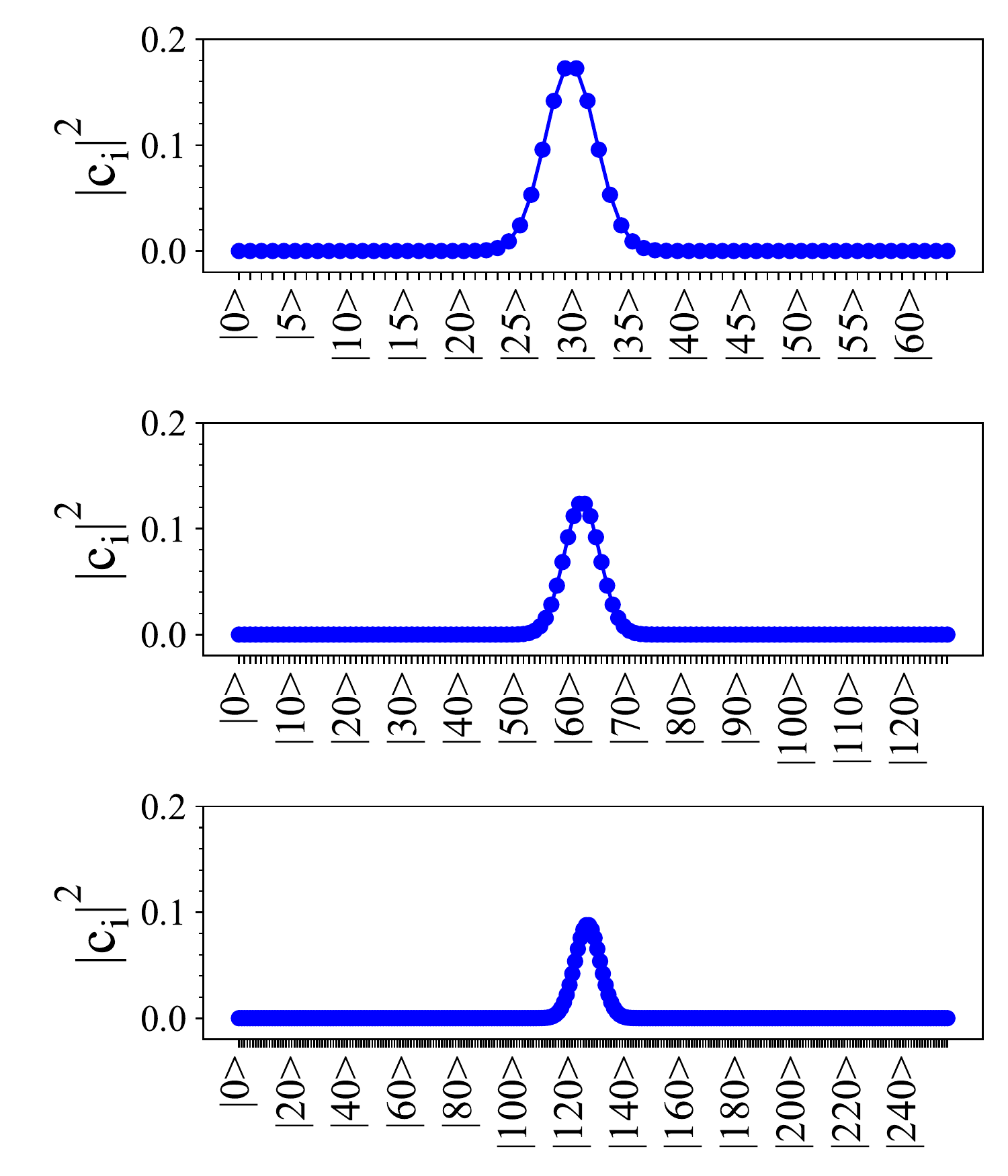} 
\caption{(Color online) An example of discretized Gaussian states in a single qudit. In this example, $N = 60$ (green), $N = 124$ (orange), $N=252$ (blue) and the bumper states are $m = 4$. Horizontal axis is the Fock states. Vertical axis is the absolute value squared of the amplitude of the field. }
\label{fig:discretizedgaussianstates}
\end{figure}

With the introduction of bumper states, the target unitary matrix $U_{\mathrm{target}}$ becomes a block matrix that contains the target $N\times N$ unitary matrix operation $V_{\mathrm{target}}$ and the block identity matrix
\begin{eqnarray}
\label{eq:targetunitary}
U_{\mathrm{target}}
=
\left(
\begin{array}{c|c}
 V_{\mathrm{target}} & 0 \\ \hline
 0 & \textbf{1}_m
\end{array}
\right).
\end{eqnarray}
Thus unitary operation on a state $|\psi\rangle$ in $N+m$ Fock states is defined as $
|\phi\rangle = U_{\mathrm{target}}|\psi\rangle,
$
where $|\phi\rangle$ is the initial state, $|\psi\rangle$ is the target state. 
We used a gradient based algorithm to find the variational parameters where the details are presented elsewhere \cite{fourierincoming}. 
Example states that represent a harmonic oscillator ground state wavefunction with $N = 60, 124, 252$ logical state and $m = 4$ bumper states are shown in Fig.~\ref{fig:discretizedgaussianstates}.

\begin{figure}
\includegraphics[width=0.45\textwidth]{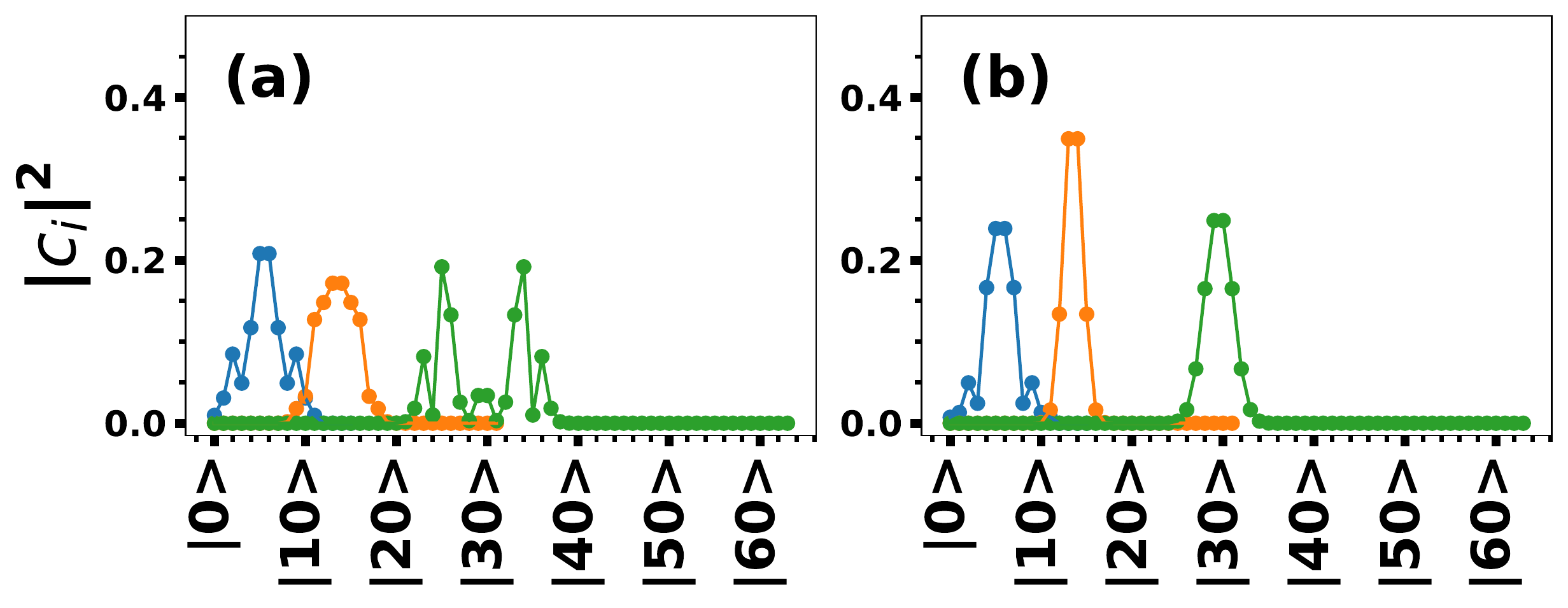} 
\includegraphics[width=0.45\textwidth]{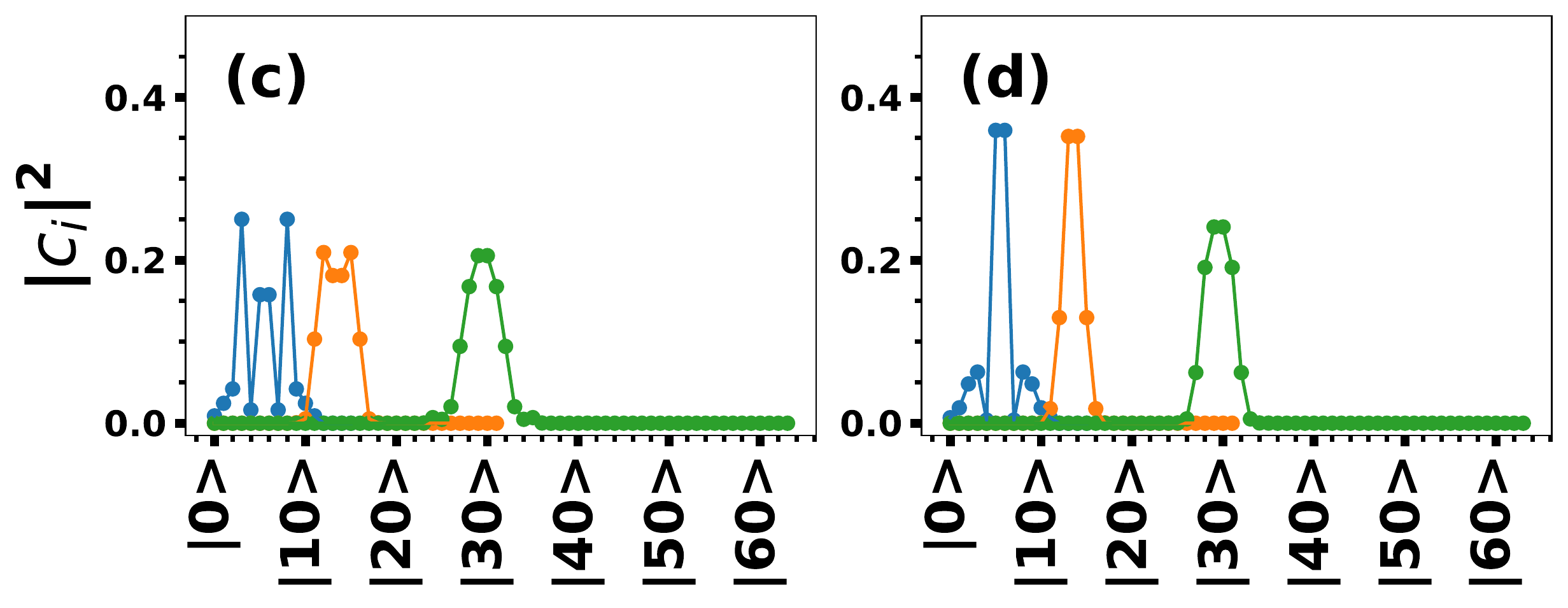}
\includegraphics[width=0.45\textwidth]{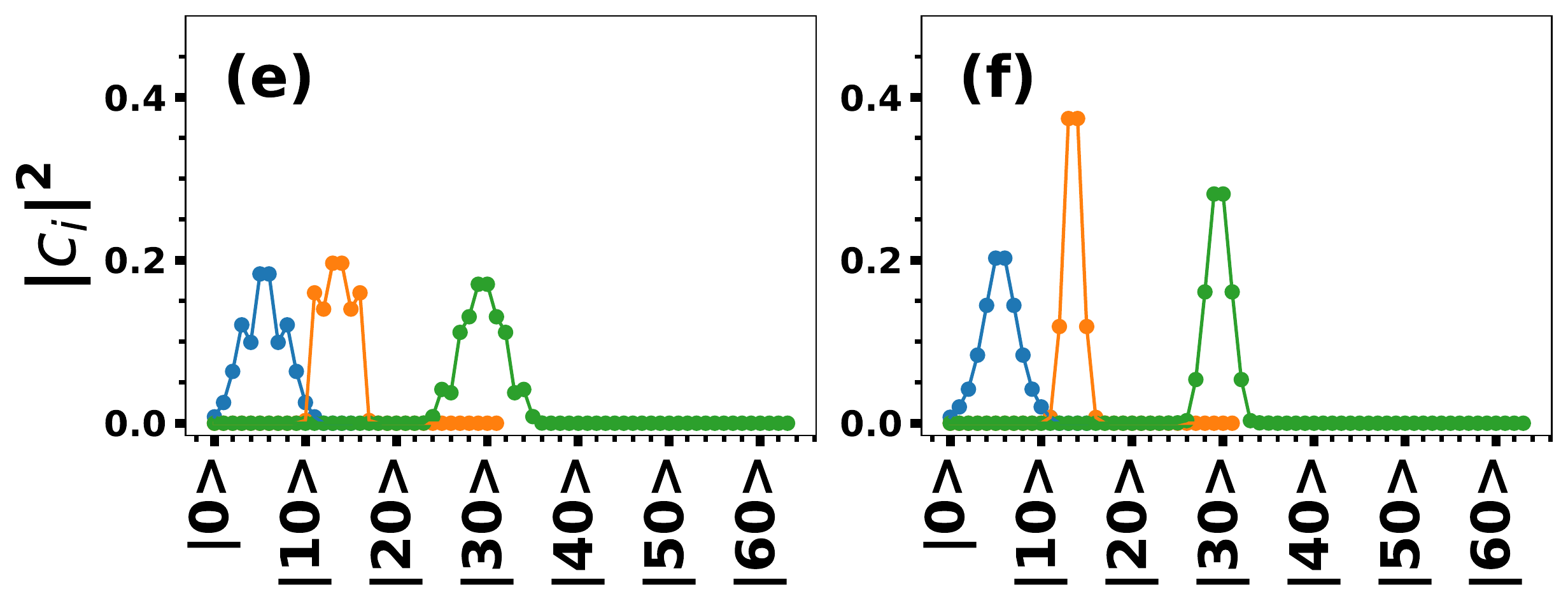} 
\caption{The ground state of a field for a single qudit with different discretizations ($N = 16$ (blue), $N = 32$ (orange), $N = 64$ (green)) with $m = 4$ bumper states are used.
The dimensionless coupling constants are 
$g = 0.5$ (a, b),
$g = 1.5$ (c, d),
$g = 2.5$ (e, f),
Left hand panel (a, c, e) are for imaginary mass, right hand panel (b, d, f) are for real mass.} 
\label{fig:fig2}
\end{figure}

The cost function we will use to prepare a single-qudit target gate is
\begin{eqnarray}
\label{eq:costfuncforgate}
\mathcal{L}_{g} = \left|\left(\frac{1}{N+m}\right)\mathrm{Tr}\left(U_{\mathrm{target}}^{\dagger}U(\vec{\alpha}, \vec{\Theta})\right)-1\right|^2.
\end{eqnarray}
After we prepare the harmonic oscillator ground state, we Trotter-simulate the field to find the ground state when the nonlinearity $g$ is present. 
If the total simulation time to find the ground state is $T = K \delta t$, the coupling constant $g$ is increased adiabatically from 0 over the time period $T$.
To find the ground state of a single qudit, we first apply the $\phi^4$ term with SNAP gates. 
The unitary operator $V$ for the Trotter step is 
\begin{widetext}
\begin{equation}
\label{eq:phi4thterm}
\begin{aligned}
V_{\phi^4}^{(s)} = e^{-i\frac{g_s}{4!}\phi_j^4 \delta t} =
\prod \limits_{n=0}^{N-1}|n\rangle\langle n|e^{-i \lambda_{n,s}(n - N/2)^4}\\
= S_N^{(4)}(-\vec{\Lambda}_s)
S_N^{(3)}\left({4 \choose 3}\frac{N}{2}\vec{\Lambda}_s\right)
S_N^{(2)}\left({4 \choose 2}\frac{N^2}{4}\vec{\Lambda}_s\right)
S_N^{(1)}\left({4 \choose 1}\frac{N^3}{8}\vec{\Lambda}_s\right)
S_N^{(0)}\left(\frac{N^4}{16}\vec{\Lambda}_s\right),
\end{aligned}
\end{equation}
\end{widetext}
where $\Lambda_s = \left\{\lambda_{n,s}\right\}$, $\lambda_{n,s} = (g_s \Delta^4 \delta t)/(4!)$, and $g_s = g(s/K)$, is the adiabatic coupling constant at time $s\delta t$ with an $s \in [0,K]$ integer and ${n \choose m} = n!/(m!(n-m)!)$ is the binomial coefficient.
We then apply the quadratic field evolution in the Eq.\ref{eq:quadraticterm}.
The next gate is the Fourier transform gate. 
In cQED systems, there are proposals that a Fourier gate can be naturally realized by using only two cavities which are coupled to one transmon on one side and taking advantage of the cross-Kerr term between two cavities by letting the transmon and cavity systems evolve over time \cite{chen2019quantum}. 
However the feasibility of this scenario is not clear for the multicell setup. 
Thus, we employ the SNAP and truncated displacement gates and construct variational block matrices to engineer single-qudit Fourier gate. 
We minimize the cost function defined in Eq.\ref{eq:costfuncforgate} and variationally find the $\vec{\alpha}$ and $\vec{\Theta}$ parameters. 
We then evolve $\delta t$ for the momentum Trotter step. 
\begin{eqnarray}
S_N^2(\vec{\theta}) = \prod \limits_{n=0}^{N-1}|n\rangle\langle n| e^{i \frac{1}{2}\delta t \Pi_n^2 }.
\end{eqnarray}
The $\Pi_n = (n - N/2)\Delta$ momentum operator is found by discrete fourier transform of the position $x_i = (i-N/2)\Delta$.
Finally, the $V_{\mathcal{F}}$ Fourier gate is applied. 
The algorithm presented here is repeated $K$ times until the total simulation time $T$ is reached. 
The ground states of a field for $N +m = 12+ 4, 28+ 4, 60+ 4$ are presented in Fig.\ref{fig:fig2}. 
The dimensionless coupling constant $g = 0.5, 1.5, 2.5$ are plugged into the code with positive $\mu^2$ (left panel) and negative $\mu^2$ values (right panel).

\begin{figure}
\includegraphics[width=0.45\textwidth]{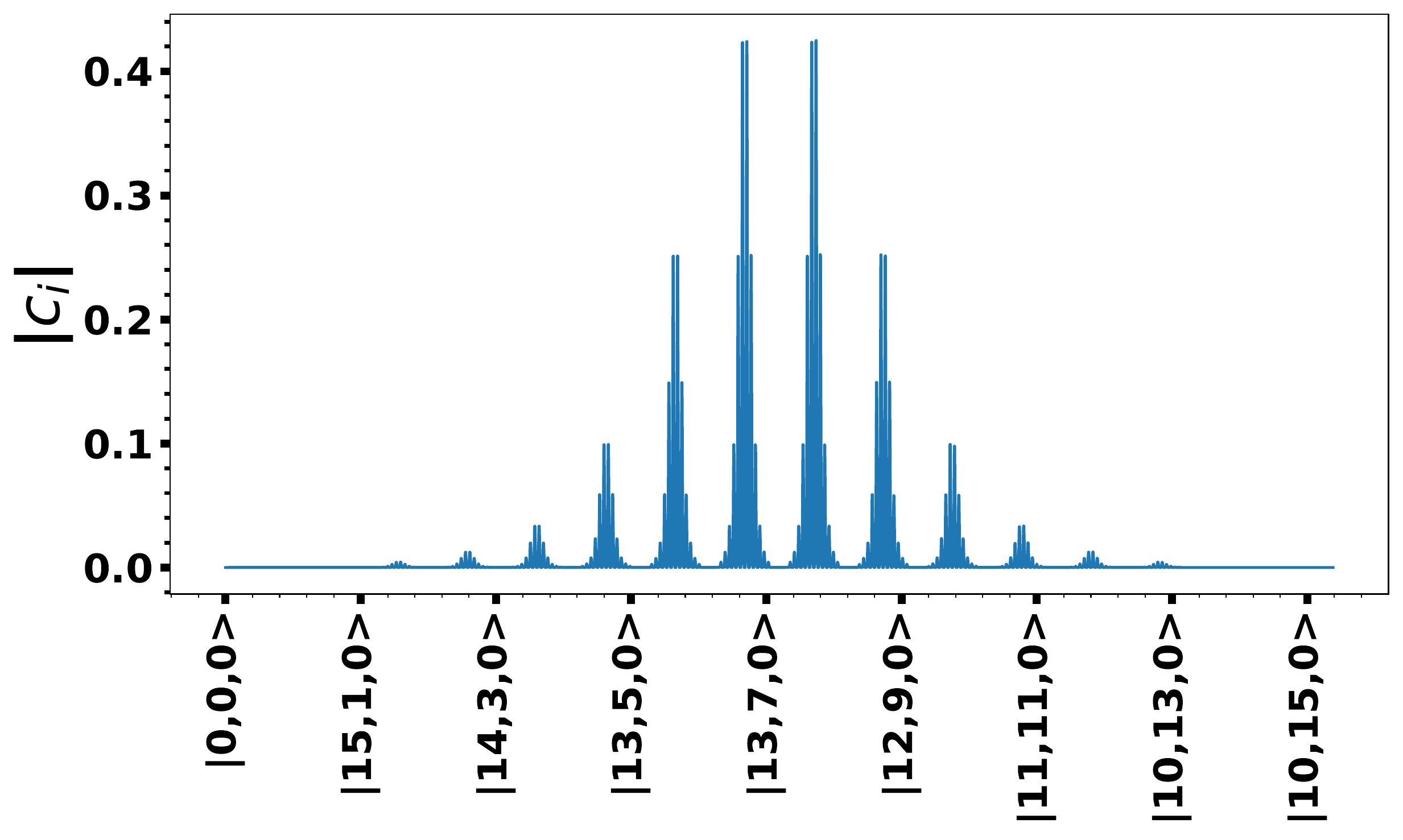} 
\label{fig:fig3}
\caption{The simulation results for three qudits and $N = 16$ with $m = 4$ bumper states. 
Coupling constants are $g = 0.5$, $f = 3.0$ and $\mu^2 \equiv -1$. 
The total simulation time in arbitrary units is $T = 2$. 
The $g$ and $f$ coupling constants are adiabatically increased through simulation time. 
The indices in the vertical axis represent the photon number in that qudit with $|q_3, q_2, q_1\rangle$. 
}
\label{fig:timesimthreequdits}
\end{figure}

%
%

\section{Multiple qudits}

In the previous chapter, we discussed how to prepare a state and a gate for a single qudit. 
A single discretized field $\phi_j$ is placed in a single qudit and ground state is found by applying phase gates over a fixed amount of time $T$.
In order to simulate more than one field, we use multiple cavities coupled to each other \cite{chakram2021seamless}.
Thus, the field discretization $j$ corresponds to the qudit index, and the position space discretization $n$ corresponds to the Fock state index in qudit $j$. 
The time simulation of a field can be realized with multicavity SNAP gate. 
The engineering of multicavity SNAP gate involves a conditional phase gate where the phase of a Fock state in a cavity mode is manipulated if a photon number on another cavity mode is satisfied \cite{chakram2021seamless, silviaprivate}. 
The experimental methods to realize SNAP and displacement gates for multiple cavities are beyond the scope of this paper.
We assume that the parameters for single SNAP gate can be used for the conditional SNAP multiple cavities by appropriate experimental techniques.
The multiqudit SNAP gate with $m$ bumper states can be constructed as 
\begin{eqnarray}
{\bf U}_{S_N}^{(k)}(\vec{\theta})_j
 = \textbf{1}_{N+m} \otimes  (...) \otimes \underbrace{ {U}_{S_N}^{(k)}(\vec{\theta})}_{jth} \otimes (...)  \otimes \textbf{1}_{N+m},
\end{eqnarray}
where 
\begin{equation}
{U}_{S_N}^{(k)}(\vec{\theta})
=
\left(
\begin{array}{c|c}
 S_N^{(k)}(\vec{\theta})  & 0 \\ \hline
 0 & \textbf{1}_m
\end{array}
\right).    
\end{equation}

We first prepare the initial multiqudit state by using the SNAP gates. 
The multi-cavity state is the tensor product of single-cavity states
\begin{eqnarray}
|\Psi\rangle = |\psi_1\rangle \otimes |\psi_2\rangle \otimes (...).
\end{eqnarray}
Once the single qudit initial state is variationally prepared with SNAP and displacement gates, the same variational parameters can be used at each qudit for conditional SNAP gates to prepare the multiqudit state. 
After this state preparation, each qudit is in the ground state of the harmonic oscillator at $t=0$. 
Next, the ground state of the field at each qudit is prepared when the interaction is present. 
The ground state preparation is made using the same algorithm we presented in the single qudit section, where we apply $V_{\phi^4}^{(s)}$, $V_{\phi^2}$, $\mathcal{F}_N$, $V_{\phi^2}$, $\mathcal{F}_N$ consecutively at each time $\delta t$ to each qudit. 

\begin{figure*}
\includegraphics[width=\textwidth]{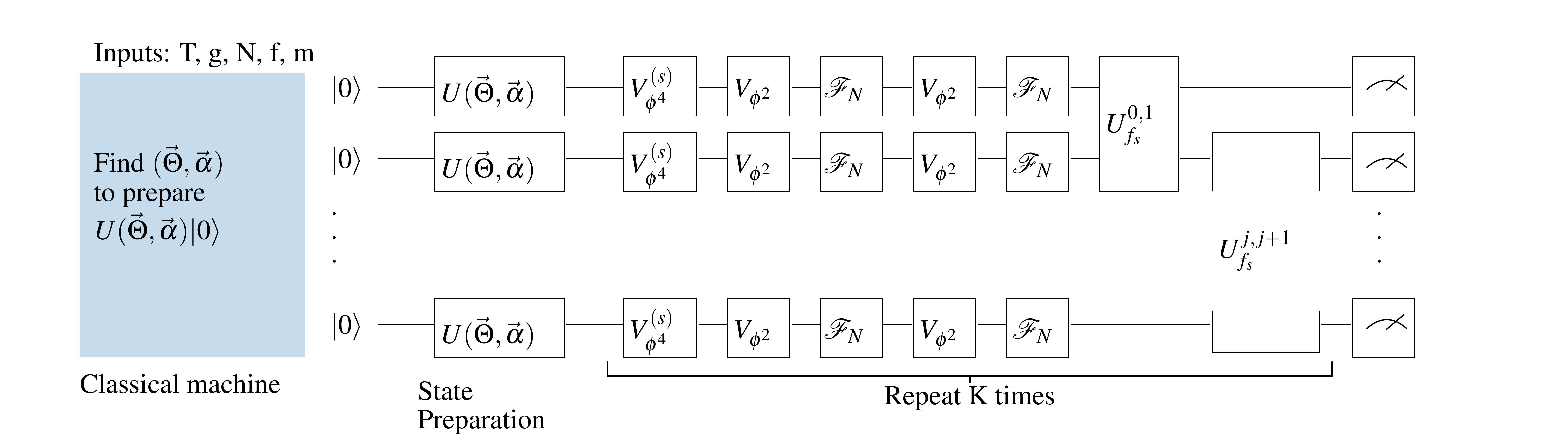} 
\label{fig:fig4}
\caption{The circuit diagram for Trotter simulation. 
First, the $\left(\vec{\Theta}, \vec{\alpha}\right)$ parameters are variationally found on a classical computer. These parameters are used on the qudit device to prepare the state. 
Then, the Trotter steps are applied to each state $K$ times and the states are measured. 
}
\end{figure*}

\section{Simulation Methods}

In this section, we discuss the time-simulation algorithm for $\phi^4$ type Hamiltonian in a multi-qudit system. 
In our simulations on a PC, three qudits are considered. 
We have already discussed engineering of the unitary gates with SNAP and displacement gate decomposition and the ground state preparation only with SNAP gates in multi-qudit systems. 
The simulation algorithm begins with the ground state $|0\rangle$ at each qudit. 
We then apply SNAP and displacement gates in order to prepare the harmonic oscillator ground state at each qudit. 
After state preparation, we first find the ground state of the interacting Hamiltonian when the coupling term $f$ is set to zero by using the algorithm we presented in the previous section.
After the ground state of an interacting Hamiltonian is found at each qudit, the $(N+m)\times(N+m)$ two qudit coupling term
\begin{eqnarray}
U_{f_s}^{j,j+1} = e^{-i f_s \Delta^2 (n_j - N/2)(n_{j+1} - N/2)\delta t},
\end{eqnarray}
where $n_j \in [0,N-1]$, is applied to two adjacent qudits.
The $f_s$ is adiabatically increased from $0$ to the final value $f$ over time $T$.

The algorithm is summarized in FIG.\ref{fig:fig4}
We provide results for $q=3$ qudits in Fig.~\ref{fig:timesimthreequdits}, simulated on a classical computer. 
The $x$ axis represent the photon number at each cavity. 
The variational parameters to realize Fourier gate are found by 
a gradient method, where the details are explained elsewhere.
The total simulation time is $T = 2 = (2000)\times \delta t$ where $\delta t = 0.001$. 
Smaller time-separations $\delta t$ of less than $0.001$ did not noticeably affect the outcome of the simulation.

\section{Conclusions}
We discussed application of the SNAP gate method in cavity systems for quantum simulation. 
Due to the fact that SNAP gates can be photon number dependent, they are excellent candidates for the simulation of field theories. 
We presented an algorithm to time-simulate a scalar field theory which has $\phi^4$ type interaction. 
Since the phases in the SNAP gates can be arbitrarily manipulated, the field theory simulation with arbitrary coupling strengths $g$ can be simulated in cavity systems.

\section{Acknowledgements}

This material is based upon work supported by the U.S. Department of Energy, Office of Science, National Quantum Information Science Research Centers, Superconducting Quantum Materials and Systems Center (SQMS) under the contract No. DE-AC02-07CH11359. 
We thank Hank Lamm, Norman Tubman for their helpful comments. 

\bibliographystyle{apsrev4-2}
\bibliography{phi4th}

\end{document}